\documentclass[useAMS,usenatbib]{mnras}
\usepackage{graphicx}
\usepackage{tikz}
\usepackage{amsfonts} 
\usepackage{xcolor}
\usepackage{mathabx} 
\usepackage[T1]{fontenc}

\title[Galactic interstellar filaments as probed by LOFAR and \textit{Planck}]{Galactic interstellar filaments as probed by LOFAR and \textit{Planck}}

\author[Zaroubi et al.]{S. Zaroubi$^1$\thanks{E-mail:
saleem@astro.rug.nl}, V. Jeli\'c$^{1,2,3}$, A.~G. de Bruyn$^{1,2}$, F. Boulanger$^4$, A. Bracco$^{5,4}$, R. Kooistra$^{1}$,\newauthor M.~I.~R. Alves$^{6,7}$, M.~A. Brentjens$^{2}$, K. Ferri\`ere$^{6,7}$, T. Ghosh$^{4}$, L.~V.~E. Koopmans$^{1}$, \newauthor F. Levrier$^8$, M.-A. Miville-Desch\^enes$^4$, L. Montier$^{6,7}$, V.~N. Pandey$^{2}$, J. D. Soler$^4$
\\ $^1$Kapteyn Astronomical Institute, University of Groningen, Landleven 12, 9747AD Groningen, the Netherlands.
\\ $^2$ASTRON, The Netherlands Institute for Radio Astronomy, PO Box 2, 7990 AA Dwingeloo, the Netherlands. 
\\ $^3$Ru{\dj}er Bo\v{s}kovi\'{c} Institute, Bijeni\v{c}ka cesta 54, 10000 Zagreb, Croatia
\\ $^4$Institut d'Astrophysique Spatiale, CNRS (UMR8617) Universit\'e Paris-Sud 11, B\^atiment 121, Orsay, France
\\ $^5$Laboratoire d'Astrophysique (AIM) Paris-Saclay, CEA Saclay, 91191 Gif-sur-Yvette
\\ $^6$CNRS, IRAP, 9 Av. colonel Roche, BP 44346, F-31028 Toulouse cedex 4, France
\\ $^7$Universit\'e de Toulouse, UPS-OMP, IRAP, F-31028 Toulouse cedex 4, France
\\ $^8$LERMA, Observatoire de Paris, PSL Research University, CNRS, Sorbonne Universit\'es, UPMC, ENS, F-75005, Paris, France
}

\begin{document}

\maketitle

\begin{abstract}
Recent Low Frequency Array (LOFAR) observations at $115-175\,{\rm MHz}$ of a field at medium Galactic latitudes (centered at the bright quasar 3C196) have shown striking filamentary structures in polarization that extend over more than 4 degrees across the sky. In addition, the \textit{Planck} satellite has released full sky maps of the dust emission in polarization at 353\,GHz. The LOFAR data resolve Faraday structures along the line of sight, whereas the \textit{Planck} dust polarization maps probe the orientation of the sky projected magnetic field component. Hence, no apparent correlation between the two is expected.  Here we report a surprising, yet clear, correlation  between the filamentary structures, detected with LOFAR, and the magnetic field orientation, probed by the \textit{Planck} satellite. This finding points to a common, yet unclear, physical origin of the two measurements in this specific area in the sky. A number of follow-up multi-frequency studies are proposed to shed  light on this unexpected finding.
\end{abstract}

\begin{keywords}
ISM: general, magnetic fields, structure -- radio continuum: general, ISM -- submillimetre: ISM -- techniques: interferometric, polarimetric 
\end{keywords}

\section{Introduction}
\label{sec:intro}

Non-thermal plasma is one of the main components of the interstellar medium (ISM) in our Galaxy \citep{ferriere01}.  It mostly contains relativistic electrons and protons, which permeate the ISM and interact with the all-pervasive magnetic field of our Galaxy \citep{haverkorn15}, resulting in synchrotron emission. At low radio frequencies, like those measured by the LOFAR telescope \citep{haarlem13}, the line-of-sight (LOS) magnetic field component ($B_\parallel$) and ionized gas can be probed by changes in the polarization of the Galactic synchrotron radiation through the so-called Faraday rotation. 

Recent WSRT and LOFAR observations \citep{bernardi09, iacobelli13, jelic14, jelic15} have shown surprisingsly strong polarization at frequencies around 150\,MHz. This polarization originates in the diffuse Galactic background. Due to Faraday rotation in the  medium between source and observer the frequency dependence of the polarization can be very  complex. The frequency-dependent polarization can be unravelled using a technique called rotation measure (RM) synthesis \citep{brentjens05}; this produces images as a function of RM or Faraday depth.  The observed structures in Faraday space can be very complex morphologically. In particular, the 3C196 field \citep{jelic15} shows a strikingly straight filamentary structure that extends over scales of more than 4 degrees. This filament also exhibits narrow depolarization canals, with a width down to $1~{\rm arcmin}$, that extend along the whole structure \citep{jelic15}.

At sub-millimetre wavelengths, in particular at the frequency band of $353$\,GHz, the \textit{Planck} satellite measured dust polarization with unprecedented sensitivity \citep{planck2015-XIX, planck2015-a01}. Polarization results from the alignment of interstellar dust grains with respect to the Galactic magnetic field \citep{davis51, lazarian07}. Hence, dust polarized emission provides powerful insights into the physical properties of the dust and the magnetic field.

The physical quantities that the two probes measure are in general thought to be   unrelated. Whereas Faraday rotation at low frequencies is sensitive to localized structures in Faraday space, which in the 3C196 field mostly have small Faraday depth, the \textit{Planck} dust polarization gives the density weighted mean orientation of the magnetic field component  perpendicular to the LOS, $\left<\mathbf{B}_\perp\right>$. Indeed, in a number of LOFAR observational windows, e.g., ELAIS-N1 and NCP \citep{jelic14}, no correlation between the two probes is evident. However,  in this letter, we report a strong correlation between the two probes in the 3C196 field. We discuss possible further observations that might help understand the physical origin of this correlation.

In \S~\ref{sec:lofar} and \S~\ref{sec:planck} of this letter,  the LOFAR and \textit{Planck} satellite data towards the 3C196 quasar are presented, respectively. In \S~\ref{sec:correlation} we discuss the correlation between the two fields and we conclude with discussion and summary in \S~\ref{sec:summary}.

\section{The LOFAR 3C196 data}
\label{sec:lofar}

The 3C196 field is one of the primary fields of the LOFAR-Epoch of Reionization key science project (de Bruyn, in prep.), centered on the very bright compact quasar 3C196. It is located  in a radio cold region towards the Galactic anticentre ($l=173^\circ$, $b=+33^\circ$). It has been observed on a large number of nights with the LOFAR High Band Antennas (HBA). For this paper we use images based on one fully processed 8h synthesis, taken in December 2012. A detailed description of the data, their processing and the image quality obtained from them are found in \citet{jelic15}. For completeness, we present here a brief overview of the data.

The dataset contains all four correlation products between pairs of orthogonal LOFAR dipoles from which full Stokes (IQUV) images were made in the frequency range of $115-175\,{\rm MHz}$ with $183\,{\rm kHz}$ resolution. The images were produced with a frequency independent resolution of about 3 arcminutes. On these images we apply RM synthesis \citep{brentjens05} to study the linearly polarized emission as a function of Faraday depth. Faraday depth gives the amount of Faraday rotation due to the intervening magneto-ionic medium between the source of the polarized emission and the observer, 
\begin{equation}
\Phi=0.81\,\mathrm{rad\, m^{-2}} \int_{\mathrm{source}}^{\mathrm{observer}} n_e B_\parallel \mathrm{d} l.
\label{eq:Faraday}
\end{equation}
Here $n_e$ is the electron number density in cm$^{-3}$, $B_{\parallel}$ is measured in $\mu$G, and $\mathrm{d}l$ is the LOS element in pc. The spectral resolution and frequency range of the data constrain the resolution of our cubes in Faraday depth to $\delta\Phi=0.9\,{\rm rad~m^{-2}}$ and the largest Faraday structure that can be detected $\Delta\Phi=1.1\,{\rm rad~m^{-2}}$ \citep{brentjens05}.  

\begin{figure}
\centering
\includegraphics[width=.45\textwidth]{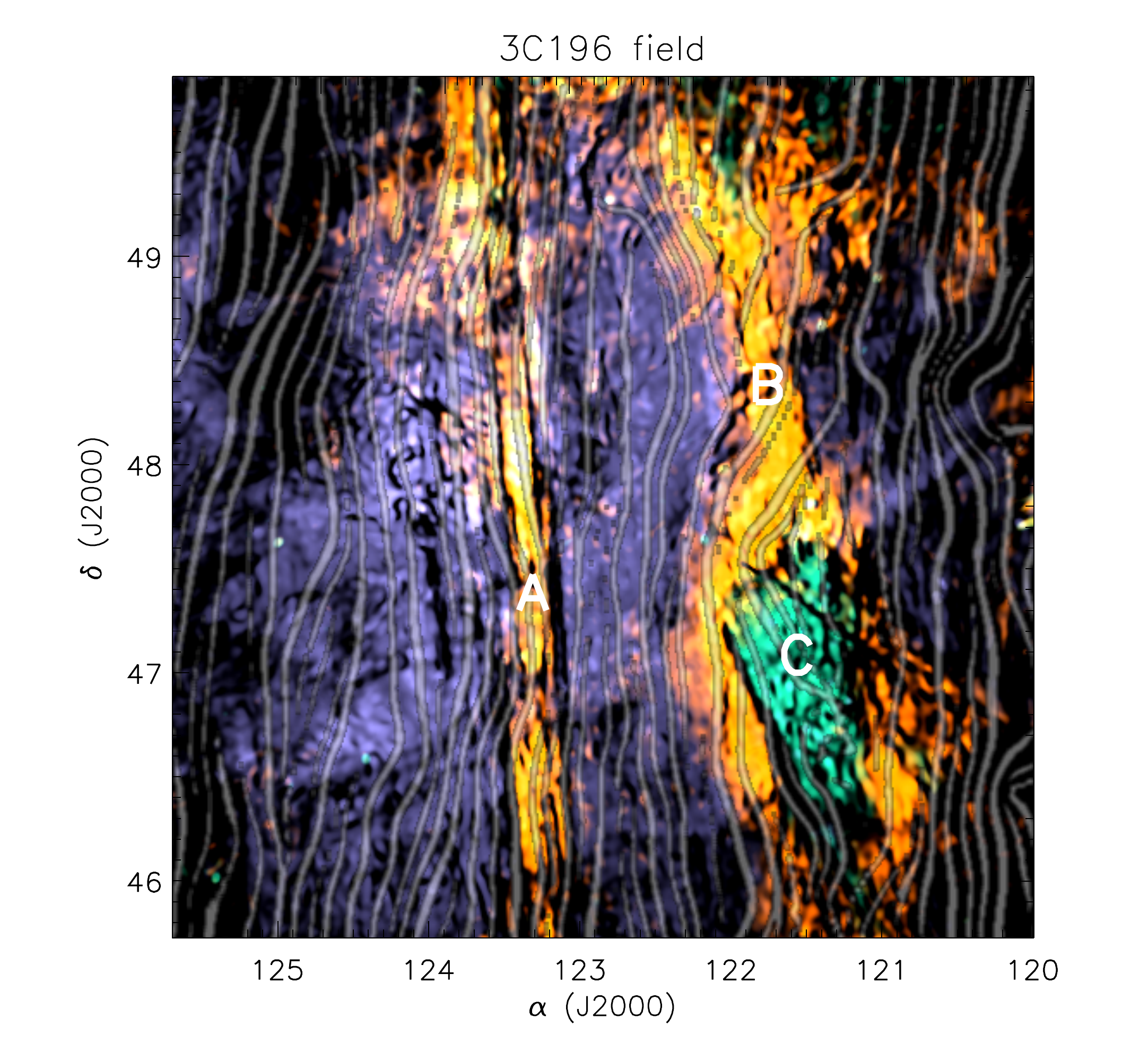}
\caption{A composite image of morphological features of the 3C196 field detected with LOFAR at different Faraday depths and the magnetic field lines orientation (gray lines) inferred from the \textit{Planck} dust polarization maps at $353$\,GHz. A ``triangular'' feature displayed in green (marked with C) is emission at negative Faraday depths ($-3$ to $-0.5\,{\rm rad~m^{-2}}$), the filamentary structures at Faraday depth of $+0.5\,{\rm rad~m^{-2}}$ are given in yellow (marked with A and B). The violet  shows the prominent diffuse background emission arising at Faraday depths from +1.0 to $+4.5\,{\rm rad~m^{-2}}$. The resolution of the LOFAR image is $3\,{\rm arcmin}$.}
\label{fig:filaments}
\end{figure}

Detected structures of Galactic polarized emission are spread over a wide range of Faraday depths (from $-3$ to $+8\,{\rm rad~m^{-2}}$) and their peak brightness temperature is between $5$ and $15\,{\rm K}$ \citep{jelic15}. Figure~\ref{fig:filaments} shows the most interesting morphological features of this emission in a composite image of polarized structures at different Faraday depths. A ``triangular'' feature showing emission at negative Faraday depths ($-3$ to $-0.5\,{\rm rad~m^{-2}}$) is shown in green. Relatively straight filamentary structures running from South to North at Faraday depths around $+0.5\,{\rm rad~m^{-2}}$ are shown in yellow, while the prominent diffuse background emission arising at Faraday depths from +1.0 to $+4.5\,{\rm rad~m^{-2}}$ is given in violet. The filaments are displaced by $\approx -1.5\,{\rm rad~m^{-2}}$ in Faraday depth relative to the surrounding background emission, suggesting a magneto-ionic medium located in front of the bulk of the emission \citep{jelic15}. Note that Faraday depth is uncertain at the level of $\approx0.1\,{\rm rad~m^{-2}}$ due to the ionospheric correction uncertainty.

\section{The \textit{Planck} magnetic field data}
\label{sec:planck}
The \textit{Planck} satellite has measured the sky in 9 frequencies ranging from $30$\,GHz to $857$\,GHz. An inspection of these maps reveals that the lowest frequencies ($30$\,GHz and $70$\,GHz) and the highest frequencies ($\ge 217$\,GHz) are dominated by Galactic synchrotron emission and thermal dust emission, respectively \citep{planck2015-a01}. In particular, the \textit{Planck} satellite team used the polarized data at $\nu = 353$\,GHz to construct an all-sky map of the Galactic magnetic field orientation projected onto the plane of the sky \citep{planck2015-a01}. 

Interstellar dust particles are non-spherical, which gives rise to polarized emission reflecting this asphericity. Coherent polarized emission is observed because the elongated dust grains tend to be aligned, along their short axes, with respect to the magnetic field. The details of the grain alignment process are not completely clear but there is a broad agreement that radiative torques are an effective means of grain alignment in the diffuse ISM \citep[for more details see][]{draine96,lazarian07} are an effective means of the observed grain alignment in the diffuse medium \citep{planck2015-XX}. As a result, the dust polarization angle is at $90^\circ$ of the perpendicular component of the magnetic field vector averaged along the LOS, $\left<\mathbf{B}_\perp\right>$. 

For this work, we use the line integral convolution \citep[LIC;][]{cabral93} representation of the magnetic field inferred from the 353\,GHz polarization maps publicly available at the \textit{Planck} Legacy Archive\footnote{http://pla.esac.esa.int}, smoothed to $15$~{arcmin} resolution. LIC is a representation algorithm of a pseudo-vector field obtained by computing the stream line that starts at the centre of each pixel. For this study, we use the LIC map corresponding to a maximum integrated length of one fourth of the map size per iteration and three iterations of the algorithm. These LIC magnetic field lines are shown in Fig.~\ref{fig:filaments}. The paper results are robust with respect to the specific choice of LIC algorithm parameters.

\section{Comparison between the maps and its interpretation}
\label{sec:correlation}

We now inspect the various correlated features between the two data sets. There are three regions of interest, marked as \textbf{A}, \textbf{B} and \textbf{C} in Fig.~\ref{fig:filaments}. The strength of the correlation between the two probes is discussed separately for each region. In general, there is a broad alignment between the South to North orientation of the magnetic field lines as measured by \textit{Planck} with the orientation of the LOFAR filaments detected in Faraday depth.

\begin{figure} \centering
\includegraphics[width=.45\textwidth]{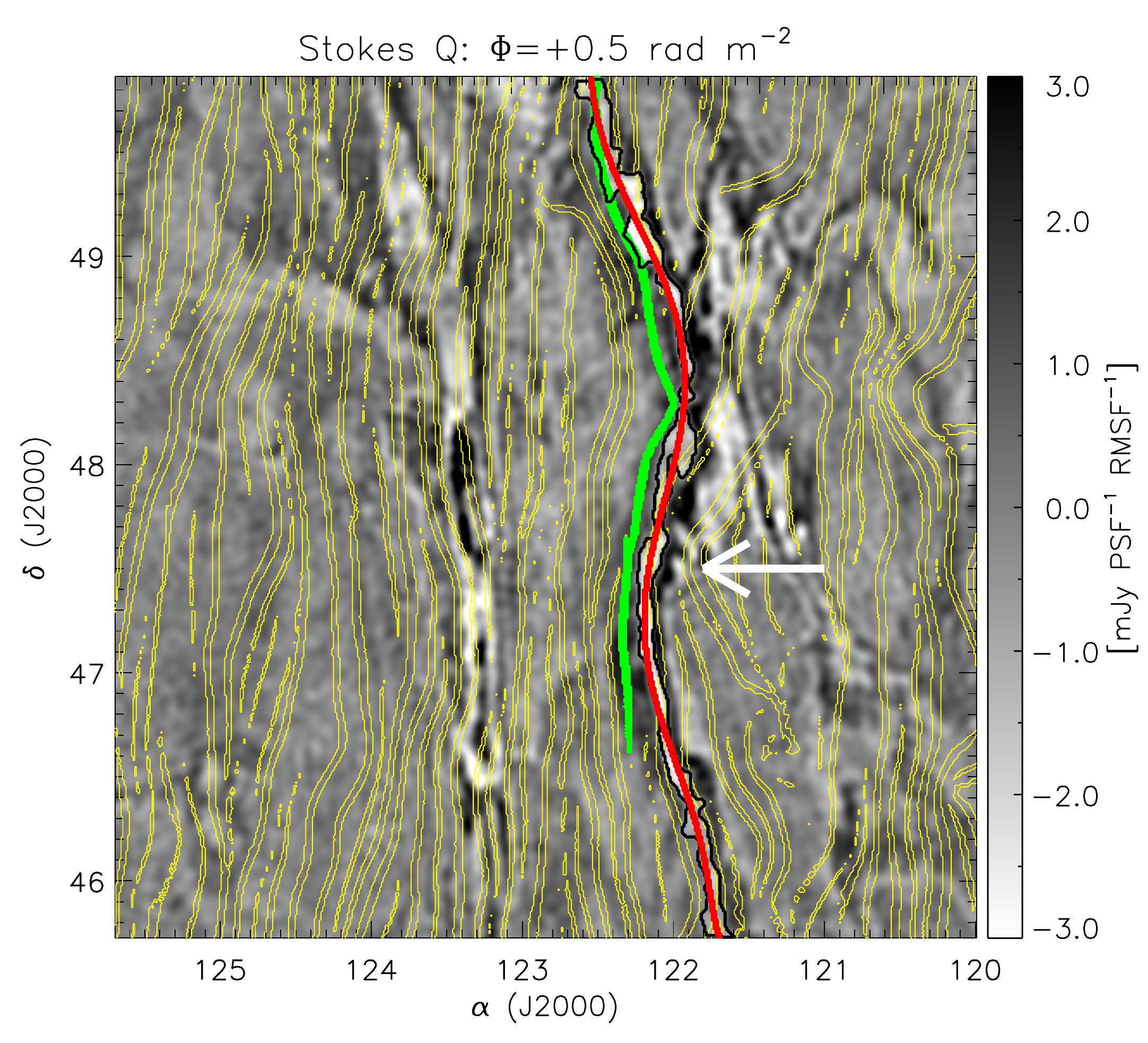}
\caption{This figure shows in gray scale the Stokes Q  image at Faraday depth of $+0.5~{\rm rad~m^{-2}}$ \citep[see][for details]{jelic15}. The area highlighted with the black contour is the region in which the polarization angle is roughly uniform. The red line shows an order 10 polynomial fit to the black contour. The \textit{Planck} magnetic field lines are overlaid and indicated with the open yellow contours. The magnetic field line shown in green marks the line that has a shape similar to the line of uniform polarization angle.}
\label{fig:Qlines}
\end{figure}

In particular, the linearly polarized emission at Faraday depth of $+0.5\,{\rm rad~m^{-2}}$ at the centre of the image (shown in yellow and marked as \textbf{A}) is very strongly aligned with the \textit{Planck} magnetic field orientation. This alignment is not only with the feature at this specific Faraday depth in the LOFAR data but also with many other features associated with this filament, such as the depolarization canals described in \citet{jelic15}. This correlation might not be surprising since the overall Galactic magnetic field direction, is aligned with the South-North direction in this specific projection. However, we note that such a correlation is not seen in the other LOFAR observational windows, at comparable Galactic latitudes like the ELAIS-N1 \citep{jelic14} and NCP (Jeli\'c et al. in prep.) fields. 

The second correlated feature is shown in the region \textbf{B}.  Here the  bending of the  magnetic field line follows that of the filament at the Faraday depth of $+0.5\,{\rm rad~m^{-2}}$. The correlation of this feature in Fig.~\ref{fig:filaments}, though obvious, requires a more careful statistical analysis to assess its significance. In order to quantify this we resort to the RM cubes. We use Stokes Q images that contain information about both intensity and polarization angle and therefore exhibit larger contrast, which helps defining the features at Faraday depth of $+0.5\,{\rm rad~m^{-2}}$ (see Fig.~\ref{fig:Qlines}). The area deliniated with a black contour is the region in which the polarization angle is roughly constant \citep[see][for more details]{jelic15}. The \textit{Planck} magnetic field lines are overlaid and indicated with the yellow contours. The filled line in green  marks the \textit{Planck} magnetic field line that has a shape similar to the LOFAR feature.

\begin{figure} \centering
\includegraphics[width=.45\textwidth]{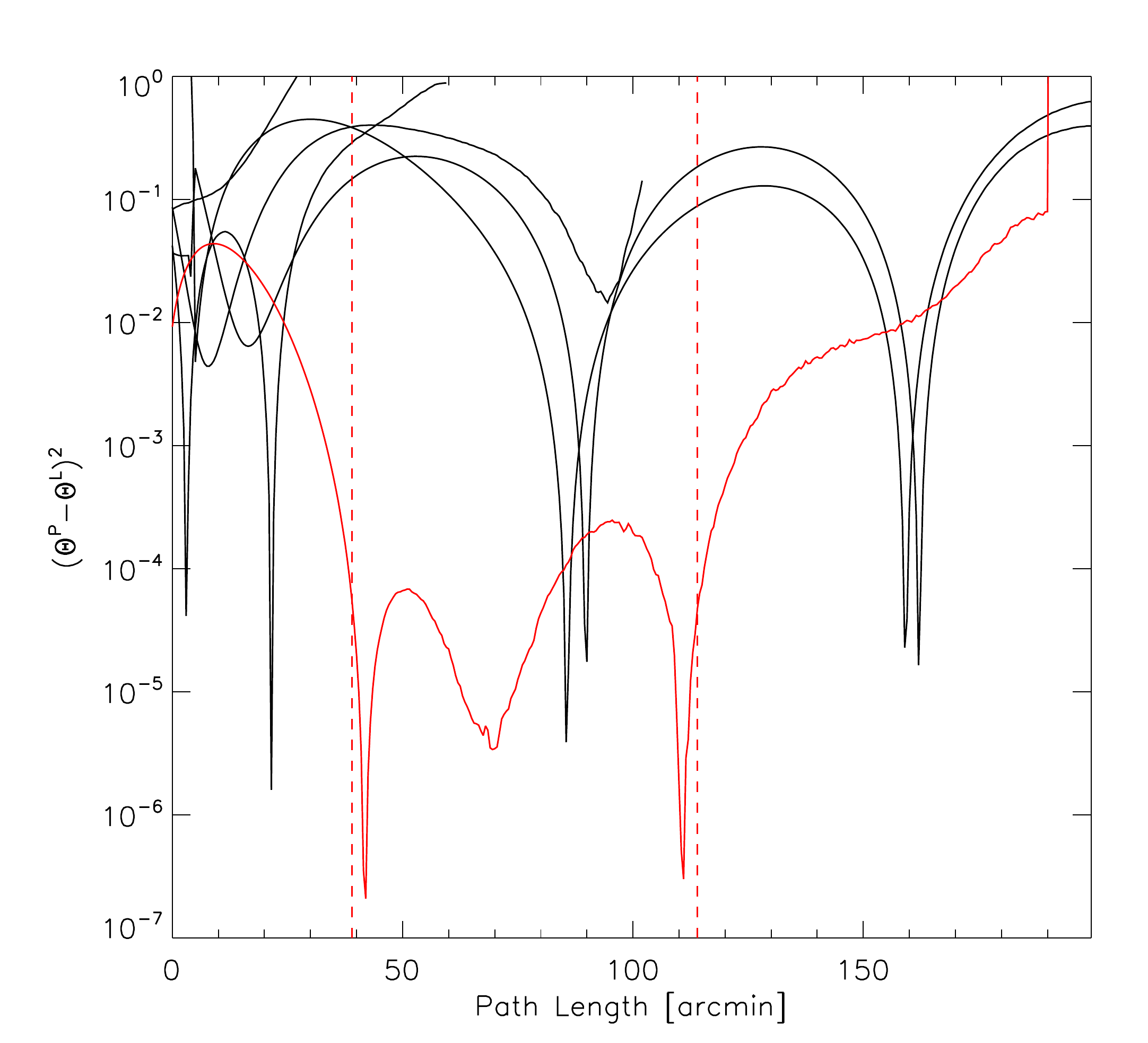}
\caption{
This figure shows the square of the difference between the angle of the tangent at each point along the LOFAR and \textit{Planck} curves, $\theta^L$ and $\theta^P$, respectively. The red curve in the figure shows the squared difference between the angles of the most similar lines in Figure~\ref{fig:Qlines} (red and green lines), whereas the rest shows the angle difference squared between the yellow line and five randomly chosen \textit{Planck} magnetic field lines, also from  Figure~\ref{fig:Qlines}. The two vertical dashed lines mark the range over which the red line is the smallest; namely $\lesssim 10^{-4}$ for about $75~{\rm arcmin}$).}
\label{fig:diff}
\end{figure}

In order to define these features both are fitted with order 10 polynomials, resulting in the red line for the LOFAR data shown in the figure\footnote{The choice to fit order 10 polynomial is somewhat arbitrary. It serves only to fit a smooth line to the data that allows a clear determination of the orientation of the magnetic and constant polarization angle lines. The only requirement is that the order of the polynomial should neither be too small nor too large, so as not to underfit or overfit the data, respectuvely.}. The angle, $\theta$, of the tangent at each point along the curve is calculated for both LOFAR (red) and \textit{Planck} (green) lines  and compared using a normalized $\chi$-squared statistic. For two lines, or part of them,  to be similar the angles defining their direction should be close over a long stretch of their path. Figure~\ref{fig:diff} shows the square of the difference between these angles, which we call $\theta^L$ and $\theta^P$ for LOFAR and \textit{Planck}, respectively. The red curve in Fig.~\ref{fig:diff} corresponds to the squared angle difference between the most similar lines in Fig.~\ref{fig:Qlines} (red and green lines), whereas the others correspond to the squared angle difference between the red  line and five randomly chosen \textit{Planck} magnetic field lines, also from  Fig.~\ref{fig:Qlines}. Clearly, the red curve shows the smallest difference for the longest stretch of length along the two curves. The two vertical dashed lines indicate  the range over which the squared angle difference is the smallest ($\lesssim 10^{-4}$ for about $75~{\rm arcmin}$).

Therefore, we define our statistic to be
\begin{equation}
\chi^2=\frac{1}{N_{pix}}\sum_{i \in N_{pix}} (\theta^L_i-\theta^P_i)^2,
\label{eq:chi2}
\end{equation}
where $N_{pix}$ is the number of adjacent pixels around the pixel with the minimum angle difference; $\theta^L_i$ is the angle of the LOFAR curve tangent at pixel $i$ and  $\theta^P_i$ is that of \textit{Planck}. $N_{pix}$ is chosen to be 150 pixels which corresponds to $75~{\rm arcmin}$, defined by the two dashed lines in Figure~\ref{fig:diff}. We note that since the two curves have in general different lengths, the comparison is carried out within the overlap region. In case this overlap covers less than 150 pixels, we choose $N_{pix} = \min\{150,N_{overlap}\}$, where $N_{overlap}$ is the number of pixels in the overlapping region. 

\begin{figure} \centering
\includegraphics[width=.45\textwidth]{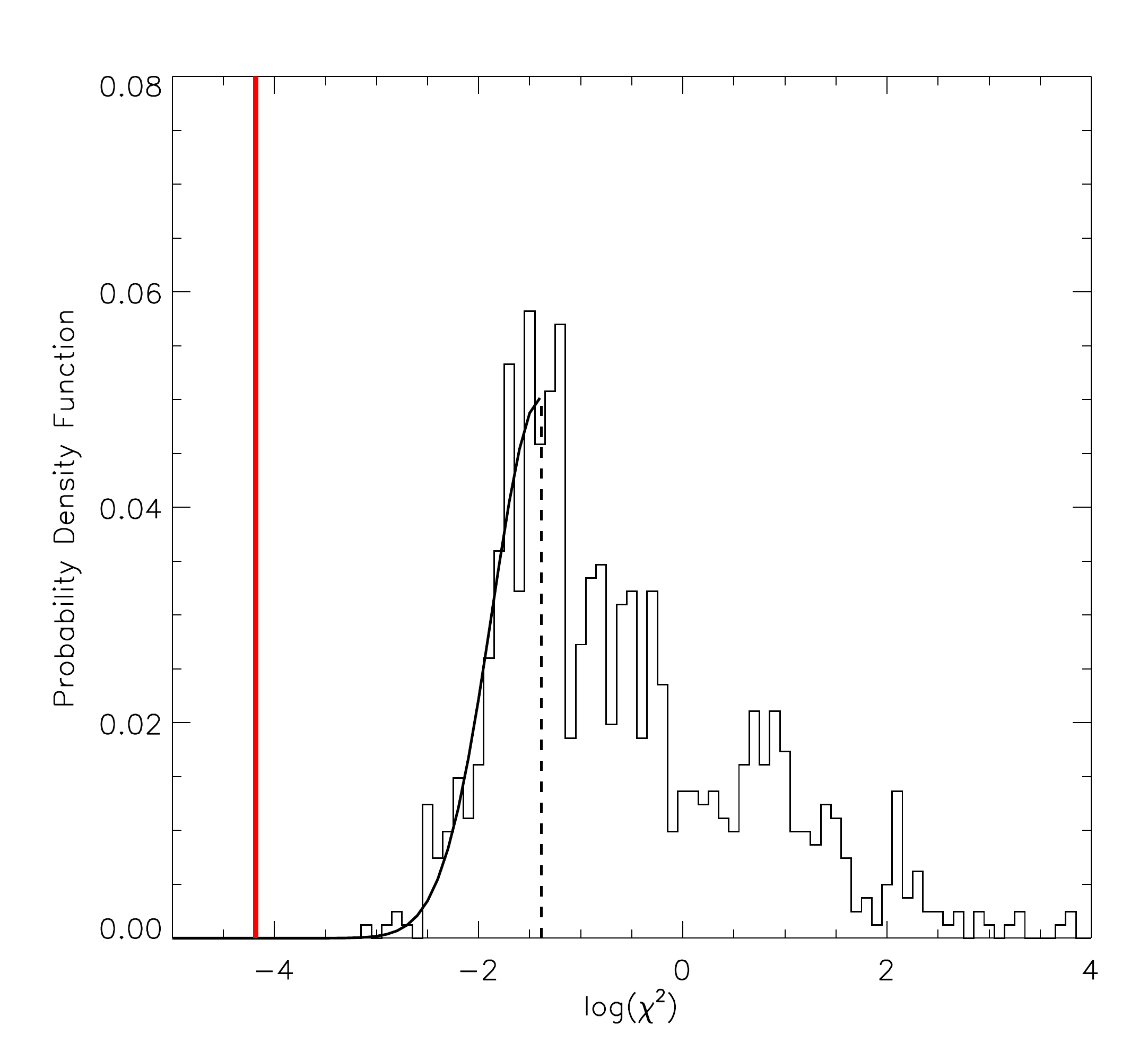}
\caption{This figure shows the probability density function of statistic defined in Eq.~\ref{eq:chi2} with logarithmic bins. The left side of the histogram is fitted with a Gaussian (the solid line). This is in practice a log-normal distribution. The Gaussian is centered at -1.38 and its width is $\sigma_{\rm G}=0.5$. The red vertical line shows the value of the $\chi^2$ statistic comparing the most similar lines in Figure~\ref{fig:Qlines} (red and green line).}
\label{fig:histogram}
\end{figure}

To address how unusually similar these two curves are, all the \textit{Planck} magnetic field lines apparent in Fig.~\ref{fig:Qlines} (about 40) are compared to the LOFAR structure of region \textbf{B} using the statistic defined in Equation~\ref{eq:chi2}. To obtain more random realizations of the magnetic field lines that retain their statistical and physical properties we randomly shift the \textit{Planck} field lines along the Declination, and if needed, we wrap them across the Declination boundaries. This results in about 1000 comparisons. The probability density function of this statistic is shown in Figure~\ref{fig:histogram} with logarithmic bins. The left side of the histogram is fitted with a Gaussian (the solid smooth line). As expected the PDF is asymmetric around its peak, however, we note that the existence of a clear sharp peak is due to the special nature of the 3C196 field in which most features are along the South-North direction. The red vertical line in this figure shows the result of the $\chi^2$ statistic for the most similar lines in Fig.~\ref{fig:Qlines} (red and green lines). This clearly shows that the similarity between these two lines is statistically significant and unlikely due to statistical coincidence. Notice that this statistic does not take into account the location of each curve, but rather compares only their shape. Therefore, the fact that these two curves are unusually similar and are co-located on the sky makes the statistical significance of this correlation even larger. We note that this result was confirmed by similar analysis based directly on the \textit{Planck} Q and U maps, independently of the LIC algorithm. 

The third region that shows correlated features is region \textbf{C} in Figure~\ref{fig:filaments}. Around this region the \textit{Planck} magnetic field lines broadly follow the bend in the red line as indicated by the arrow in Figure~\ref{fig:Qlines}. Moreover, the magnetic field lines from \textit{Planck} seem to be less coherent within this region. We have deconvolved the magnetic field into rotational (curl) and potential (divergence) components, assuming the third direction to be the LOS direction. This analysis shows that the magnetic field within region \textbf{C} have a dominant rotational component. This component can not be due to turbulence simply because region \textbf{C} has a coherent Faraday displacement. The last two statements are hard to quantify statistically, hence they should be viewed merely as potentially interesting observations.

\section{Discussion and Conclusions}
\label{sec:summary}

In the previous section we have discussed the morphology and significance of the correlated features between the LOFAR and \textit{Planck} data sets in the 3C196 field. The LOFAR polarization data in Faraday depth is proportional to $ n_e B_{\parallel}\Delta l$ (Eq.~\ref{eq:Faraday}), whereas \textit{Planck} measures, through the polarized dust emission, $\left<\mathbf{B}_\perp\right>$. Since dust and electron distributions are probably unrelated and  the magnetic field components that these probes measure are different, one does not generally expect the two data sets to be correlated. Indeed, in a number of common observational fields, we confirm that such a correlation is not evident. 
 
The correlation features reported in this paper point to a common underlying physical structure that influences both the LOFAR and \textit{Planck} data sets. One can roughly divide the features into global and local features. The global feature is the broad correlation of the dust inferred orientation of the magnetic field and the LOFAR filaments, clearly apparent in the images. This is especially striking when one inspects features \textbf{A} and \textbf{B} in Figure~\ref{fig:filaments}. \citet{jelic15} make the point that since in the 3C196 field the magnetic field that follows the spiral arm of our Galaxy is almost perpendicular to the LOS,  $B_{\parallel}$ is small relative to $\langle B_\perp \rangle$. This conclusion is supported by the estimated $\langle B_\parallel \rangle  =0.3\pm 0.1\, \mu\mathrm{G}$ in the direction of the pulsar J081558+461155, located about 2$^{\circ}$ south of 3C196, using the RM to dispersion measure ratio \citep[see][and references therein]{jelic15}. A low $B_\parallel$ suggests that the origin of the correlation resides in the alignment of the electron density distribution with $\left<\mathbf{B}_\perp\right>$. Such alignment has been observed for the filamentary structures traced by dust emission \citep{clark14, planck2015-XXXII}. In other words, the  broad alignment between the two images might be driven by the global magnetic field of the Galaxy, which is typically $\sim 5\,\mu\mathrm{G}$ \citep{haverkorn15}. This interpretation, although the most straightforward for us, does not preclude other explanations that rely on variations in the structure of $B_\parallel$.

The correlations implied by the local features are apparent in regions \textbf{B} and \textbf{C}. The agreement between the bending of the lines in both regions, especially in \textbf{B}, argues that the two effects come from the same physical location. The physical properties of the ISM in this region are uncertain. However, some constraints can be derived from the LOFAR observations \citep{jelic15}. The lack of observed emission in the LOFAR total intensity, associated with the filamentary structures detected in polarization, gives an upper limit to their thermal free-free emission, $T_{\rm ff}\lesssim 0.2$\,K. If we adopt $T_e=8000\,{\rm K}$ for the electron temperature \citep{reynolds91}, which is a typical value for the warm ionized ISM, and $\Delta l=1\,{\rm pc}$ for the LOS thickness \citep{jelic15}, we find $n_e\lesssim1\,{\rm cm^{-3}}$. The thickness of the structures in Faraday depth, $\Delta \Phi \sim 1\,\mathrm{rad\,m^{-2}}$, constrains  $B_\parallel \gtrsim 1.2\,{\rm \mu G}$, assuming again $\Delta l=1\,{\rm pc}$. Motivated by the agreement in the bending of region \textbf{B}, we consider equipartition between magnetic energy and thermal energy. Using these constraints on $n_e$, we find that the strength of the total magnetic field has to be $B_{\mathrm{total}} \lesssim 6.5\,{\rm \mu G}$. Note that in reality these values might be different as equipartition might not hold between the magentic energy and thermal energy but rather between the magnetic and kinetic energy associated with bulk motions. We also note that the true values of $B_\parallel$ and $B_{\mathrm{total}}$ are probably close to the limits derived above, because obviously $B_\parallel \leq B_{\mathrm{total}}$, moreover the detection of polarized dust emission by \textit{Planck} requires a significant $\left<\mathbf{B}_\perp\right>$. This is also consistent with the analysis of the LOFAR data in this field as noted by \citet{jelic15}.

To resolve the ambiguity of the physical parameters and shed light on the physical picture of the ISM, additional observational probes are needed. Here we mention a few. 
\begin{itemize}
\item Dust particles are usually embedded within neutral hydrogen (H{\sc i}) gas. This is especially evident through H{\sc i}/dust correlation at high Galactic latitudes, where the sky is largely transparent to H{\sc i} emission and infrared emission from dust. Therefore, H{\sc i} data can provide kinematic information about the features studied in this paper.

\item Observed starlight polarization, caused by selective absorption by magnetically aligned interstellar dust grains along the LOS, traces the orientation of $\left<\mathbf{B}_\perp\right>$ \citep{hiltner56,serkowski75, pavel14}. By repeating this measurement for stars at different distances, one can map the the 3D distribution of the orientation of $\left<\mathbf{B}_\perp\right>$ and provide complementary information to \textit{Planck}'s dust polarization observations. However, this type of polarimetry is sensitive only to the orientation of the interstellar magnetic field. 
It is worth noting that a 3D mapping of the Galactic magnetic field might be crucial for the removal of dust polarisation foreground in order to measure the primordial CMB B-mode polarisation signal \cite[see e.g.,][]{tassis15}.

\item To determine the strength of the interstellar magnetic field, one can attempt to measure the Zeeman splitting of the H{\sc i} 21-cm line or other radio lines. Such data measure the $B_\parallel$ component of interstellar magnetic field and complement observations of the Faraday depth structure.

\item Future limits on the fine structure in total intensity of the diffuse emission will allows us to constrain $\vert\mathbf{B}_\perp\vert^2$.
\end{itemize}

Finally, the correlation between these two probes calls for exploration of a similar correlation in other fields. Specifically, we propose to explore regions in which the \textit{Planck} derived magnetic field is coherent and ordered over a large region of the sky and follow such regions with LOFAR observations. This comparison might shed more light on the physical origin of this correlation.

\section*{acknowledgements}
SZ and VJ thank the Netherlands Foundation for Scientific Research  support through the VICI grant 639.043.006 and the VENI grant 639.041.336, respectively.  AGdB, FB, and LVEK acknowledge support by the European Research Council for projects 339743 (LOFARCORE), 267934 (MISTIC), and 258942 (FIRSTLIGHT), respectively. 

\bibliographystyle{mnras}

\end{document}